\documentclass[runningheads]{llncs}
\usepackage{graphicx}

\usepackage{array,ragged2e}
\usepackage{subfig}
\usepackage{float}
\begin{document}

\title{Measuring scientific buzz}

% \title{Understanding the dynamics of Artificial Intelligence research}
\author{Kishore Vasan \and Jevin West}
\authorrunning{Vasan and West.}
\institute{Information School, University of Washington, WA 98195, USA
\email{\{kishorev,jevinw\}@uw.edu}}
\maketitle              % typeset the header of the contribution
\begin{abstract}
Keywords are important for information retrieval. They are used to classify and sort papers. However, these terms can also be used to study trends within and across fields. We want to explore the lifecycle of new keywords. How often do new terms come into existence and how long till they fade out? In this paper, we present our preliminary analysis where we measure the burstiness of keywords within the field of AI. We examine 150k keywords in approximately 100k journal and conference papers.  We find that nearly 80\% of the keywords die off before year one for both journals and conferences but that terms last longer in journals versus conferences. We also observe time periods of thematic bursts in AI -- one where the terms are more neuroscience inspired and one more oriented to computational optimization. This work shows promise of using author keywords to better understand dynamics of buzz within science.

\keywords{keyword analysis \and burst detection \and survival analysis \and scientometrics \and science of science}
\end{abstract}

\section{Introduction}

Keywords can do more than just classify papers for information retrieval. They represent unique concepts associated with a paper and can be used as a proxy for knowledge creation. They can provide clues to the movement of ideas and trends within and across disciplines. For example, we can track hot terms like 'big data' to see where they originate, what disciplines they spread to and how long they last within the literature. Evaluating how terms change over time can give us insights into how disciplines evolve and respond to new trends in technology and methods. This kind of information could be useful to researchers trying to capture the pulse of a field or help them avoid 'buzzy' terms and instead focus on growing topics. Funding agencies would also find this useful for allocating funds to topics on the rise. 

In this preliminary poster, we bring together methods, not brought together before, for measuring the burstiness of keywords within the field of Artificial Intelligence (AI)\footnote{We plan to extend this to other fields and over longer time periods.} and lay groundwork for doing this more broadly. AI is known for its booms and busts. In fact, researchers often point to AI "winters". This boom and bust cycle make it ideal for studying the trendiness of jargon in the literature. We talk about the lessons learned and how these methods can be applied to other fields outside AI.

\section{Methods}
\subsection{Data}

Abstract data for this work comes from the Web of Science(WoS). A local copy of all WoS paper metadata resides in a MySQL database managed by the DataLab of the Information School at the University of Washington. From the data we filtered out all papers with subject\_traditional as 'Computer Science, Artificial Intelligence'. The reason for doing this is to restrict our dataset to only computer science research oriented AI papers since we want to focus on knowledge creation and exclude papers on applications of AI. We performed some simple keyword data cleaning procedure such as combining similar terms like 'neural network', 'neural-network', 'NEURAL NETWORK' into 'neural networks'. However, we did not want to extensively clean the keywords since a keyword could have different meanings depending on the context.\footnote{We plan to use novel techniques to cluster similar keywords together in future work.} Additionally, each paper also contained a document type entry that indicated the type of paper. We used 'Proceedings Paper; Meeting' as conference papers and 'Article; Article' as journal papers. Conference papers include conferences such as \textit{IEEE Conference on Computer Vision and Pattern Recognition} (CVPR) and journal papers include journals such as \textit{Neurocomputing}. Miscellaneous papers include books, biographies, and editorial letters. Due to missing historical data, we restrict our analysis to papers published between 1990 and 2016. 

\begin{table}[!h]
\setlength\belowcaptionskip{-20pt}
\caption{Descriptive Statistics}
\centering
\begin{tabular}{l@{\hskip 0.3in}|l@{\hskip 0.3in}l@{\hskip 0.3in}l@{\hskip 0.3in}l}
%l|llll}
\multicolumn{1}{c|}{} & Num & Papers w/ & Num & Keywords/  \\
\multicolumn{1}{c|}{Paper Type} & Papers & Keywords & Keywords & Paper \\ \hline 
Journal Papers & 43516 & 39760 & 84683 & 2.13 \\
Conference Papers  & 51639   & 29997  & 48691 & 1.62 \\
Misc Papers & 23854   & 11963  & 23492 &  1.96\\ \hline
All Papers & 119009 & 81720 & 156866 & 1.92
\end{tabular}
\vspace{-10mm}
\end{table}

\subsection{Identifying Keyword Bursts}

To measure burstiness of keywords we use Kleinberg's bursty algorithm [1]. This is a fairly well vetted popular algorithm used in the scientometrics community to map fading and emerging themes [2]. The resultant burst weight of a keyword from the algorithm takes into account the proportion of papers containing that keyword and provides a metric for strength of influence of that keyword in that 'bursty' time frame. For this initial work, we only considered keywords that appeared in atleast 20 papers. The distribution of terms in this threshold gives us a good representation of data to investigate the bursts. For the bursty analysis\footnote{Code can be found at www.github.com/kishorevasan/measuring-scientific-buzz}, we created a year-by-keyword matrix using 74232 papers and 2770 keywords. The severe drop in keywords is due to the 20 paper requirement. With this matrix, we perform the burst detection algorithm, enumerate the keywords and present the timeline of top bursts in \textbf{Fig 3}. 

\begin{minipage}{\linewidth}
	  \hspace{-1.4cm}
      \centering
      \begin{minipage}{0.45\linewidth}
          \begin{figure}[H]
              \includegraphics[width=\linewidth,height = 2.3in]{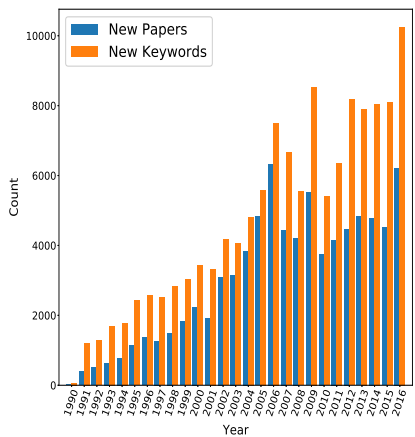}
              \caption{Creation of new papers and unique keywords over time in AI. We notice traces of AI winter with a peak of new terms in 2006, followed by another peak in 2009 and 2012.}
          \end{figure}
      \end{minipage}
      \hspace{0.05\linewidth}
      \begin{minipage}{0.45\linewidth}
          \begin{figure}[H]
              \includegraphics[width=\linewidth]{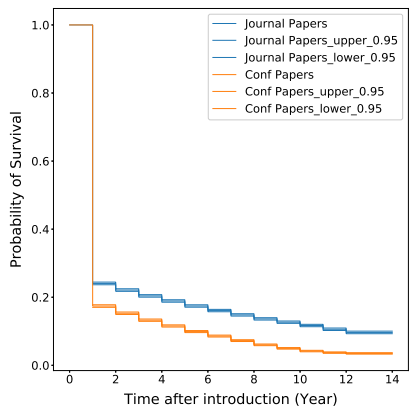}
              \caption{Kaplan-Meier survival plot of keywords. We observe that on average 80\% of keywords don't survive past year 0. We also observe a monotonic decrease in survival for both publication venues.}
          \end{figure}
      \end{minipage}
  \end{minipage}

\subsection{Survival of Keywords}

One of our main questions was to investigate how long terms last once they are newly introduced. To conduct this analysis, we restrict our dataset to keywords introduced between 2003 and 2014, thus allowing two years (2015 and 2016) for subsequent observation to see if they resurfaced. We chose this because it had good representation of terms and no major bursts in that time period. In this time frame, 38245(78.55\% of overall) new keywords were introduced by conference papers and 46252(54.61\% of overall) new keywords were introduced by journal papers. Two separate curves were fit for conference and journal papers as displayed in \textbf{Fig 2}. We used a non-parametric log rank test [4] to test if the survival curve of conference and journal keywords are identical.

\begin{figure}
\includegraphics[width=\textwidth,height = 7 cm]{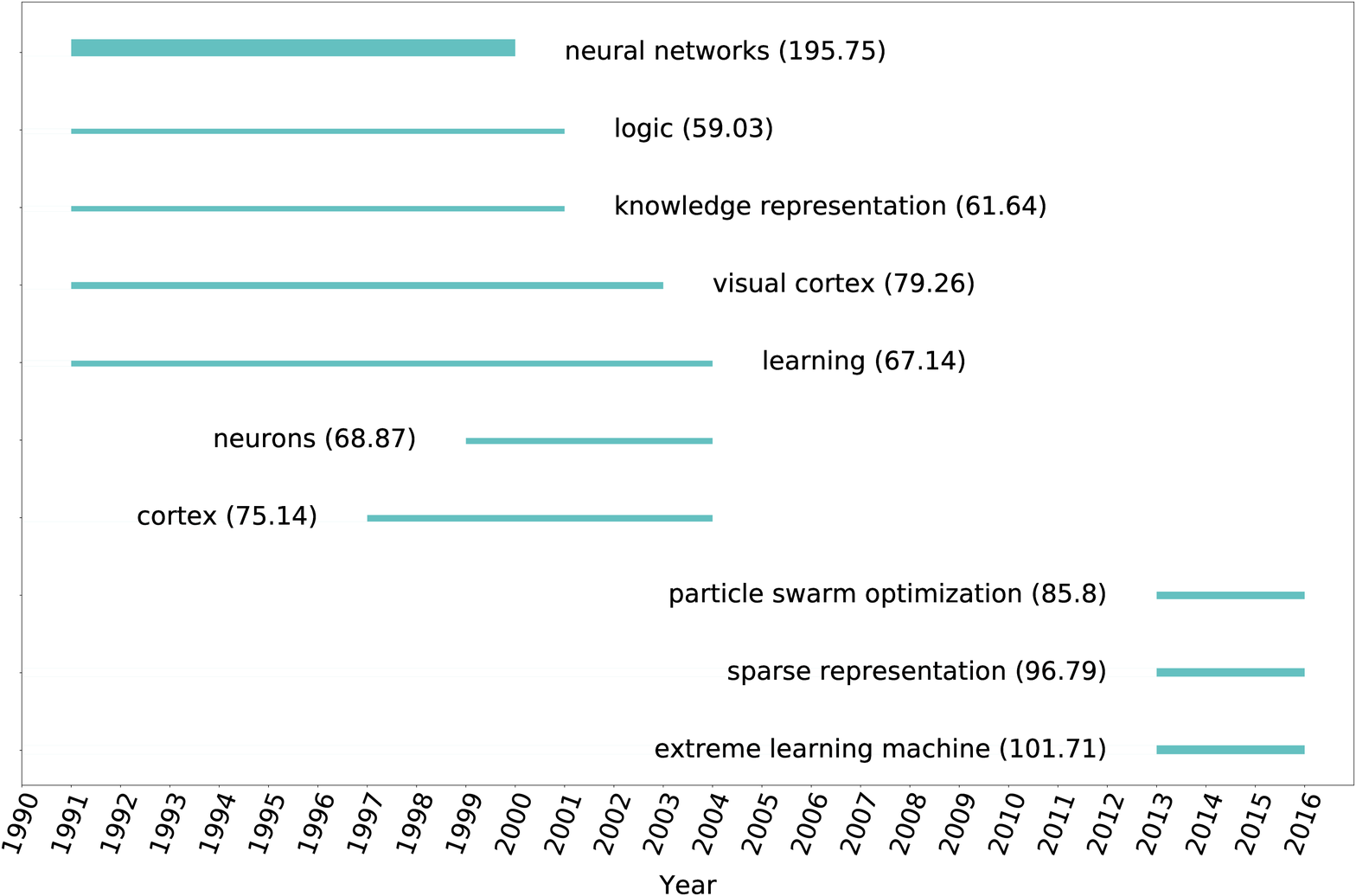}
\caption{Timeline of top 10 AI keyword bursts with bar widths scaled by strength of burst. We notice that by a huge margin, neural networks observed the biggest burst in AI from 1991 to 2000. We notice indication of AI winter with no powerful bursts between 2005 to 2012.} \label{fig2}
\vspace*{-1.3em}
\end{figure}

\section{Preliminary Results}

Keyword bursts describe thematic bursts. The major theme for keywords in the early bursts seem to be related to neuroscience ('visual cortex','neurons','cortex', 'neural networks') and the representation side of machine learning ('knowledge representation','logic','learning'), whereas recent bursts focus more on computational concepts('extreme learning machine','sparse representation','particle swarm optimization'). We notice this thematic shift over two decades that went from more neuroscience terminology, among the bursting terms, to the current 'bursty' focus on novel computational concepts. 

Another finding is that, only 8.85\% of keywords made it to more than 4 papers. This indicates that very few keywords resurface in multiple papers after introduction, at least in AI. From the log-rank test results we observe that with 0.01 level of significance, conference and journal keyword don't survive at the same rate. Keywords by journals tend to stick around longer than keywords by conferences. Publication venues are represented differently by different disciplines. We plan to look at this in other fields as well and speculate whether this is likely the case in other fields as well.

\begin{table}
\setlength\belowcaptionskip{-10pt}
  \caption{Log-Rank Test}
  \label{tab:testResults}
  \begin{tabular}{|l|c|c|c|c|c|c|}
  \hline
Group & Number & Observed(O) & Expected(E) & (O-E)\textsuperscript{2}/E & Chi Sq Test & p value\\
    \hline
    Group 1 (Journ) & 38245 &36432& 33765 & 211 & 1306 & \textless 2e-16\\ 
    Group 2 (Conf) & 46252 & 39540 &42207 & 169 & 1306 & -\\ \hline 
\end{tabular}
\vspace{-3em}
\end{table}

\section{Conclusion}
\vspace{-0.3em}
%Our goal with this poster was to begin to investigate the different ways to measure burstiness, buzziness or trendiness of keywords in hot fields like AI. We are interested in figuring out when these kinds of terms come into the literature, how long they last and when they are replaced by other bursty terms. We use keywords because of their simplicity but also because of what they convey to readers by the authors and publishers. They are one word descriptions that summarize major topic areas, which is ideal for studying buzz terms. 

Our analysis on AI keywords reveal three preliminary findings. One, most terms die out before year one. Nearly 80\% of the keywords don't make it past year zero. Two, conferences seem to have shorter-lived keywords than journals. And, three, we notice two major thematic bursts in AI. The first burst was dominated by terms that were neuronally inspired (e.g., neural network, visual cortex), while the second major burst contained computationally oriented terms (e.g., particle swarm optimization, sparse representation). We plan to extend this work to other fields in order to test how well this model identifies these major changes. 

%We found a gap in bursts between 2004 and 2013. This could be partially explained by an increase in the number of new keywords introduced in that time period (see fig 1). In this figure, the number of new unique terms peak in 2006, goes down and then rises again in 2009. It then begins to flatten in 2013. This does not fully explain why we are seeing a burst gap, but it may provide clues. Further work needs to be done to better understand this result. 

% There are limitations in what can be said about the field dynamics by looking solely at keywords. We plan to incorporate abstracts and full text when available. We also will incorporate other meta-data (titles, authors, publications) and citations. However, we do find the simplicity of keywords useful in tracking buzziness. The act of choosing single word descriptions provides a different kind of information about a paper. In aggregate, they can tell us a little bit about where a field has gone and where it is going -- and help us gain more insight into the birth of new terms and the death of well worn keywords. 


\end{document}